\documentclass[aps,prb,twocolumn,titlepage,nofootinbib]{revtex4}
\usepackage{graphicx,physics}
\usepackage{color}
\usepackage{dcolumn}
\usepackage{latexsym}

\usepackage[normalem]{ulem}
\usepackage{hyperref,amssymb}
\usepackage{url}
\usepackage{color,soul}
\usepackage{graphicx}
\usepackage{verbatim}
\usepackage{multirow}
\usepackage{amsmath}
\newcommand{\beq}{\begin{eqnarray}}
\newcommand{\eeq}{\end{eqnarray}}
\usepackage{mathrsfs}
\usepackage{float,soul}
\usepackage[dvipsnames]{xcolor}
\usepackage{mathtools}
\usepackage{slashed}
\usepackage{physics}	
\usepackage{graphicx}   
\usepackage{epstopdf}
\usepackage{subfigure}  
\usepackage{hyperref}   
\usepackage{bbold}
\usepackage{wasysym}
\usepackage{feynmp}
\usepackage{hyperref}
\hypersetup{colorlinks,
}
\usepackage{bm}

\usepackage[latin1]{inputenc}

\begin{document}

\title{Can we build a transistor using vacancy-induced bound states \\ in a topological insulator?}
\author{Cunyuan Jiang$^{1,2}$}
\email{cunyuanjiang@sjtu.edu.cn}
\address{$^1$School of Physics and Astronomy, Shanghai Jiao Tong University, Shanghai 200240, China}
\address{$^2$Wilczek Quantum Center, School of Physics and Astronomy, Shanghai Jiao Tong University, Shanghai 200240, China}

\begin{abstract}
    Topological insulators (TIs) have been considered as promising candidates for next generation of electronic devices due to their topologically protected quantum transport phenomena. In this work, a scheme for atomic-scale field effect transistor (FET) based on vacancy-induced edge states in TIs is promoted. By designing the positions of vacancies, the closed channel between source and drain terminals provided by vacancy-induced edge states can have the energy spectra with a gap between edge and bulk states. When gate terminal receive the signal, electric field applied by gate terminal can shift quasi Fermi energy of the closed channel from edge states into the gap, and hence open the channel between source and drain terminals. The energy spectra and the effect of electric field are demonstrated using Haldane model and density functional theory (DFT) respectively. This work suggest possible revolutionary applicational potentials of vacancy-induced edge states in topological insulators for atomic-scale electronics.
\end{abstract}
\maketitle

\section{Introduction}
From the end of 20th century, topological insulators (TIs), as a new class of quantum material distinguished from conventional insulators, has roused widespread attentions for both experimental and theoretical investigations.\cite{PismaZhETF.42.145,PhysRevLett.95.226801,PhysRevLett.96.106802,PANKRATOV198793,doi:10.1126/science.1148047} As its name 'insulator', the interior of TIs is electronic insulating as an ordinary insulator because of topological energy gap between valance and conduction bands, however, different from ordinary insulator, the surface of TIs is electronic conducting due to presence of edge states.\cite{PhysRevLett.117.235304,Xia2009-ep,Yan2013-us} The conducting edge states is robust against disorder,\cite{Kotta2020,Mitchell2018,PhysRevLett.118.236402} and any continue deformation with a profound topological geometry reason since the edge states they are born when a surface unties 'knotted' electron wavefunctions.\cite{Moore2010} Because of unique contribution of edges states, various transport phenomena in TIs have been reported, such as quantum anomalous Hall effect,\cite{RevModPhys.95.011002} and quantum spin Hall effect,\cite{doi:10.1126/science.1148047} topological superconductivity,\cite{RevModPhys.83.1057} and anomalous conductivity induced by disorder.\cite{PhysRevLett.129.196601} 

With these special transport properties, TIs is also considered for designing new type of electronic devices.\cite{Gilbert2021} Including topological p-n junction,\cite{PhysRevB.85.235131,Tu2016} topological field effect transistor (FET),\cite{Vandenberghe2017,10606287,doi:10.1126/science.1256815,PhysRevLett.123.206801} topological interconnect,\cite{Gupta2014} and so on. Although the robustness and dissipation-free of edge states are advantages of using TIs for electronic devices. However, the physical difficulties are still there, and feasible schemes are still needed. For example, FET, one of the most fundamental element in logic circuit, have been designed based on electric field induced phase transition between trivial phase between topological phase in a 2D thin film,\cite{doi:10.1126/science.1256815,Ezawa_2014} but the required field intensity is strong.\cite{Vandenberghe2017} The application 3D TIs is more complicated because of inevitable shunting paths through surface of the 3D device.\cite{Vandenberghe2017} In addition, another knotty challenge arises because the energy spectra of edge states typically overlap with bulk states, resulting in a fully occupied topological energy gap. The absence of an energy gap means the Fermi energy (\(E_F\)) invariably intersects with allowed states-whether edge or bulk states-making it difficult to achieve a open-circuit condition.\cite{PhysRevLett.117.235304,Xia2009-ep,Yan2013-us} These challenges of applying in electronic devices calling more detailed and specific design according to the features of TIs.

As development of nano-technology, atomic scale manipulation of defects in crystalline materials arise great attentions because of their non-trivial effects.\cite{Mao2016,doi:10.34133/2019/4641739} For TIs, the point defect, vacancy, is also reported to inducing localized edge states around and hence anomalous conductivity when the localized states connect with each other.\cite{PhysRevLett.129.196601,PhysRevLett.107.116803} Furthermore, previous work reported the dynamics of vacancies induced conducting excitations and suggested their application potential for atomic scale resistive circuit.\cite{brotherpaper} With knowing the dynamics of vacancies induced edge states in TIs, the difficulties limit applications of topological edge states can be solved and hence new opportunities for atomic scale topological electronic devices beyond resistive circuit is indicated. 

In this work, a scheme of atomic scale FET electronic device based on vacancies in TIs is proposed. The main mechanism is to create an energy gap between topological edge states and trivial bulk states, the energy gap which allows for open-circuit condition when quasi Fermi level (\(E_F^*\)) is shifted into the energy gap by electric field effect, and for closed-circuit condition when quasi Fermi level intersects with topological edge states. The gapped energy band structure will be verified by Haldane model with vacancies,\cite{PhysRevLett.61.2015} and the open-circuit condition driven by electric field will be also confirmed through density functional theory (DFT) simulation.\cite{RevModPhys.87.897,PhysRev.140.A1133} FET, resistor and conducting wire are basic components for logic circuit. Together with the previous design of atomic scale resistor and wire,\cite{brotherpaper} the scheme of atomic scale FET proposed in this work is able to be used for designing practical atomic scale logic circuits, and therefore provide special opportunities for atomic scale low power electronics and even for relative industries.

\section{Gapped band structure of vacancies chain in TIs}
\begin{figure}[ht!]
    \centering
    \includegraphics[width=\linewidth]{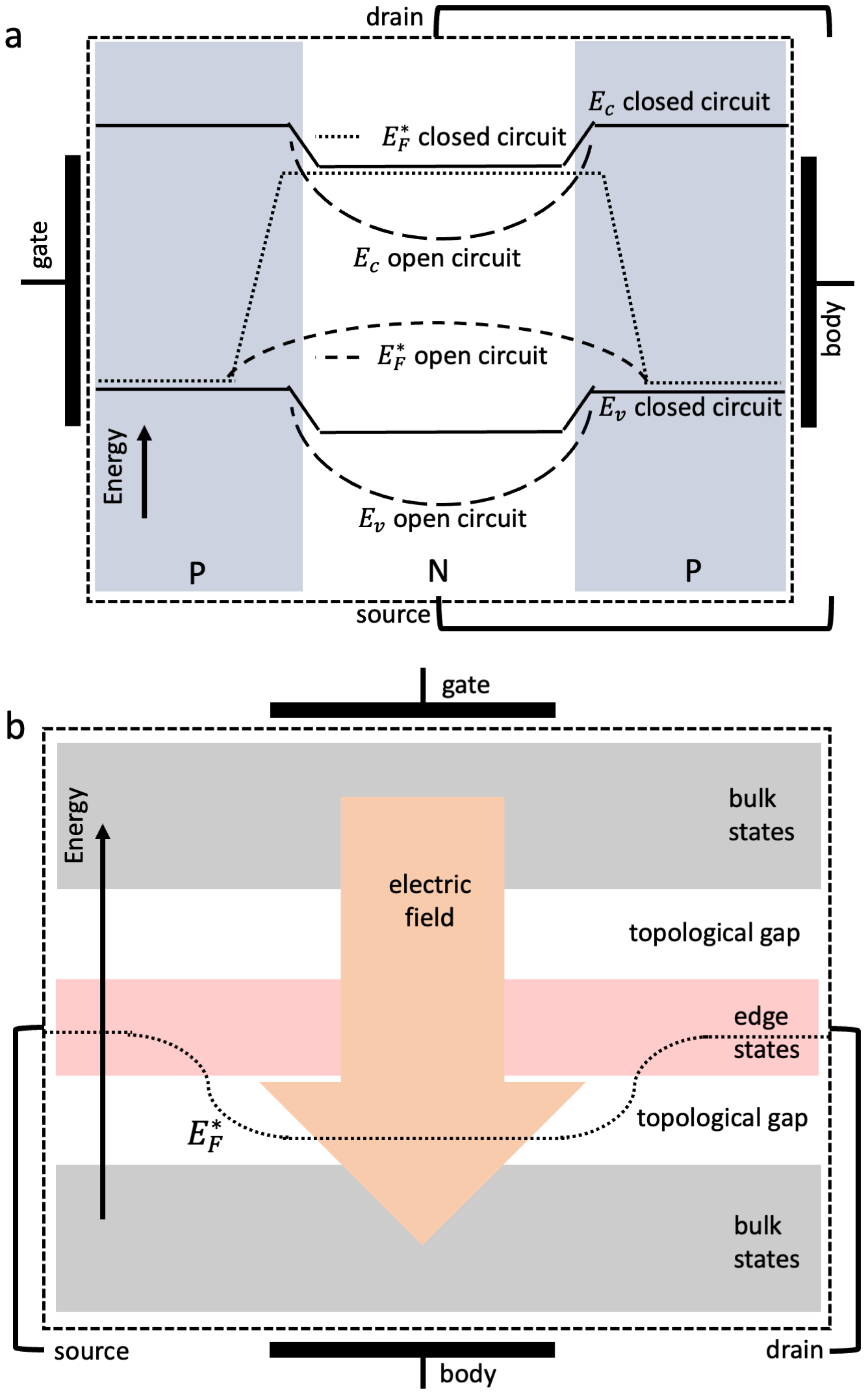}
    \caption{\textbf{a}, schematically showing the mechanism of semiconductor junction FET. Lighter blue regions are P-type semiconductor connected with gate terminal and the white region is N-type semiconductor connecting source and drain terminals. The quasi Fermi level (\(E_F^*\)) of electrons intersects with conduction band in N-type semiconductor under closed circuit condition, and lies in band gap away from conduction band under open circuit condition. \(E_c\) and \(E_v\) are the energy of conduction band and valance band respectively. \textbf{b}, the scheme of FET in TIs using topological states. The energy bands must have a topological energy gap (white area) between valance and conduction bulk states (gray area). The topological edge states (lighter red area) in topological energy gap making closed circuit between source and drain terminals when there is no voltage on gate terminal. Under the opposite situation, the electric field applied through gate terminal will shift the quasi Fermi level (\(E_F^*\)) into topological energy gap between edge states and bulk states, and hence leads to open circuit between source and drain terminals.}
    \label{figfermi}
\end{figure}

Let's start with a quick review of semiconductor junction FET.\cite{semiconductorbook} The mechanism of a semiconductor junction FET relies on controlling the flow of carriers between the source and drain terminals by modulating the width of the depletion region in the p-n junction. By applying a reverse bias to the gate terminal, the depletion region expands, constricting the conductive channel. When the depletion region fully "pinches off" the channel, carrier flow ceases, resulting in an open-circuit condition between the source and drain. The mechanism can be interpreted using the concept of quasi Fermi energy, which is defined as \(E_F^* = E_F + k_B T \mathrm{ln}(\rho / \rho_{0})\), with \(E_F\) is equilibrium Fermi energy, \(k_B\) the Boltzmann constant, \(T\) temperature, \(\rho\) carrier concentration, \(\rho_{equilibrium}\) equilibrium carrier concentration. As shown in Fig\ref{figfermi}-a, under closed circuit condition, the quasi Fermi level intersects with conduction band in N-type region and therefore allow current to pass through between source and drain. Under open circuit condition, the quasi Fermi level is not able to reach the conduction band due to expansion of depletion region, and hence current is not allowed to pass. The decrease of quasi Fermi level is effect of build-in electric field in depletion region that reduce density of carriers \(\rho\). 

Similar to semiconductor junction FET, a scheme of junction FET in TIs can also reach open circuit condition by shifting quasi Fermi level into an energy gap through field effect as shown schematically in Fig.\ref{figfermi}-b.

\begin{figure}[ht!]
    \centering
    \includegraphics[width=\linewidth]{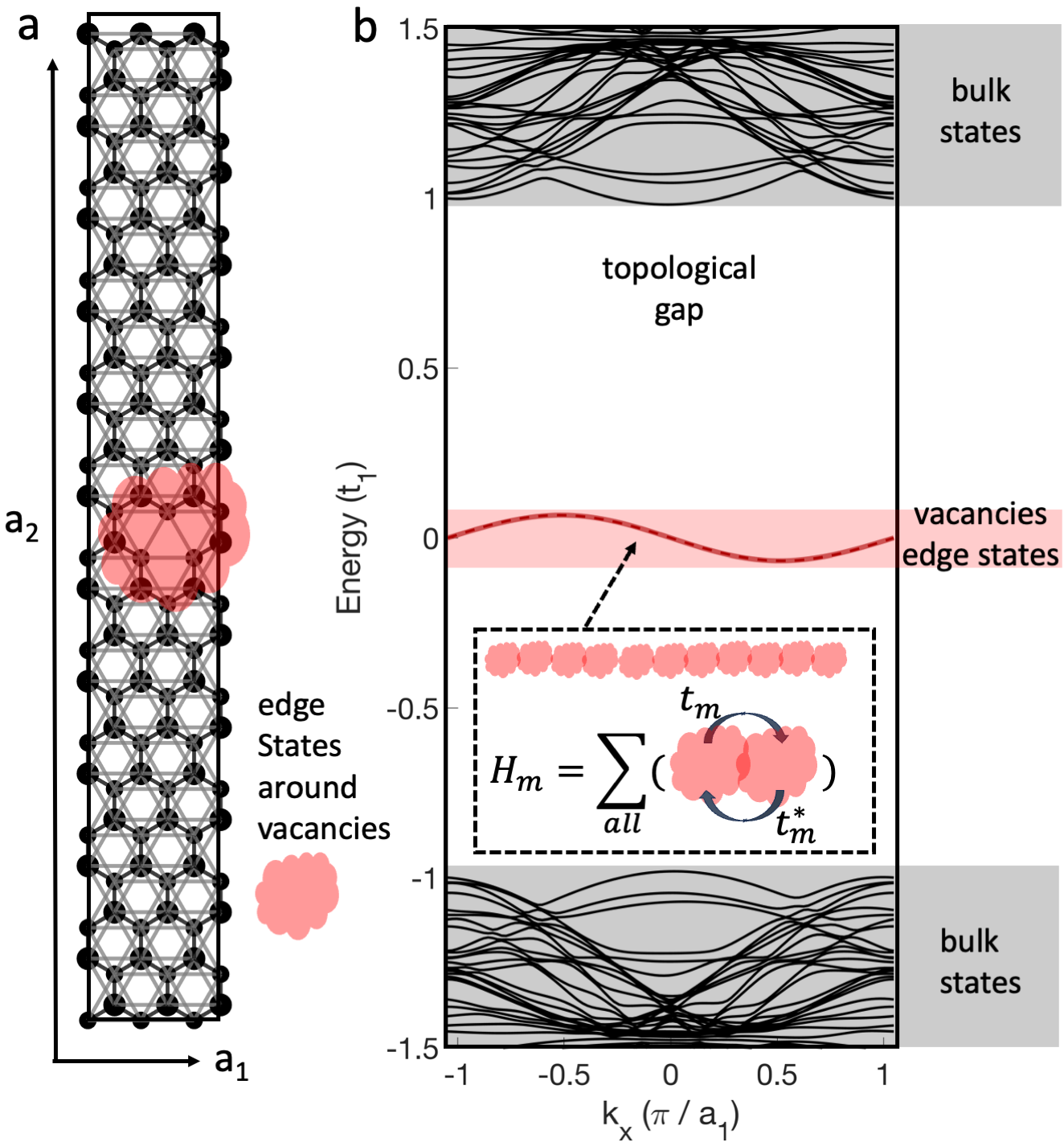}
    \caption{The structure and energy bands of a vacancies' chain. \textbf{a}, the structure of super cell of Haldane model. \(\boldsymbol{a}_1\) and \(\boldsymbol{a}_2\) are super cell lattice vectors. The big and small black dots are A and B type of atoms. Black and gray bounds between atoms indicate the nearest neighbor hopping and the next nearest hopping respectively. The lighter red cloud indicates position of vacancy and associated edge mode around. \textbf{b}, energy band structure of the super cell. Gray and lighter red area indicate bulk and edge states induced by vacancies respectively. The thin dashed line above the red topological edge band is predicted by the effective Hamiltonian in the dashed box. The effective Hamiltonian consider hopping of vacancy induced edge state to its neighbor site on a chain of vacancies with hopping rate \(t_m\).}
    \label{figband}
\end{figure}

TIs have intrinsic topological protected energy gap for bulk states as mentioned above, however the edge states typically fill the topological gap and overlap with bulk states.\cite{PhysRevLett.117.235304,Xia2009-ep,Yan2013-us} As the consequence, current using edge states is not able to reach open circuit condition, which limits the design of TIs FET. This issue can be overcame using edge states induced by vacancies in TIs, since the dynamics of such states is quantitatively under control. The effective Hamiltonian of vacancy states can be write in a tight binding form,\cite{brotherpaper}
\begin{equation}
    H_m = \sum_{i,j}  t_{m,ij} a_{j}^\dagger a_i , \label{realh}
\end{equation}
with \(a_{j}^\dagger (a_{j})\) creation (annihilation) operator of vacancy states at position \(\boldsymbol{r}_j\), and \( t_{m,ij}\) the effective hopping parameter determined by the overlap between two nearby vacancies states. This effective Hamiltonian is schematically shown in the insert on Fig.\ref{figband}-b. With the effective Hamiltonian in Eq.\eqref{realh}, it can be expected that a \(1D\) chain of vacancies, as a conducting wire, its band structure has finite range in energy. According to the effective model, the band structure of \(1D\) chain is \(E(k) = t_m \mathrm{exp}(-i k a) + t_m^* \mathrm{exp}(i k a) = - 2 |t_m| \mathrm{sin}(ka)\), with \(t_m = i |t_m|\) a pure imaginary number which is required by the nature of topological edge states as in Haldane model.\cite{PhysRevLett.61.2015} In Fig.\ref{figband}, full band structure including both edge and bulk states of \(2D\) TIs with a chain of vacancies is simulated using Haldane model,\cite{PhysRevLett.61.2015} with more details in Supplementary Information, SI-1. In the simulation, a super cell with \(3 \times 22\) unit cells of honeycomb lattice is used as shown in Fig.\ref{figband}-a. A vacancy is made in the center of such super cell. By applying periodical boundary, the vacancy is allowed to interact with another nearby vacancy along \(\boldsymbol{a}_1\) direction due to overlap of their vacancy induced edge states. The interaction along \(\boldsymbol{a}_2\) direction is not allowed because of large spacing. Such a system, TIs with conducting wire made by vacancies, show a typical band structure that edge states induced by vacancies lie in the center of topological gap and separated from bulk state as shown in Fig.\ref{figband}-b. As expected through the effective model in Eq.\eqref{realh}, the band structure is indeed a sine function.

\section{Structure of FET and the field effect}
Such a band structure that topological edge states do not fill the topological gap provide the opportunities for reaching open circuit condition by shifting quasi Fermi level \(E_F^*\) into the gap between edge states and bulk states, as shown in Fig.\ref{figfermi}-b. According to the relation between quasi Fermi level \(E_F^*\) and local electron density \(\rho\), \(E_F^* \approx E_F + k_B T \mathrm{ln}(\rho / \rho_{0})\), the quasi Fermi level can be shifted by reducing the local electron density through external electronic field from gate terminal. 

\begin{figure}[ht!]
    \centering
    \includegraphics[width=\linewidth]{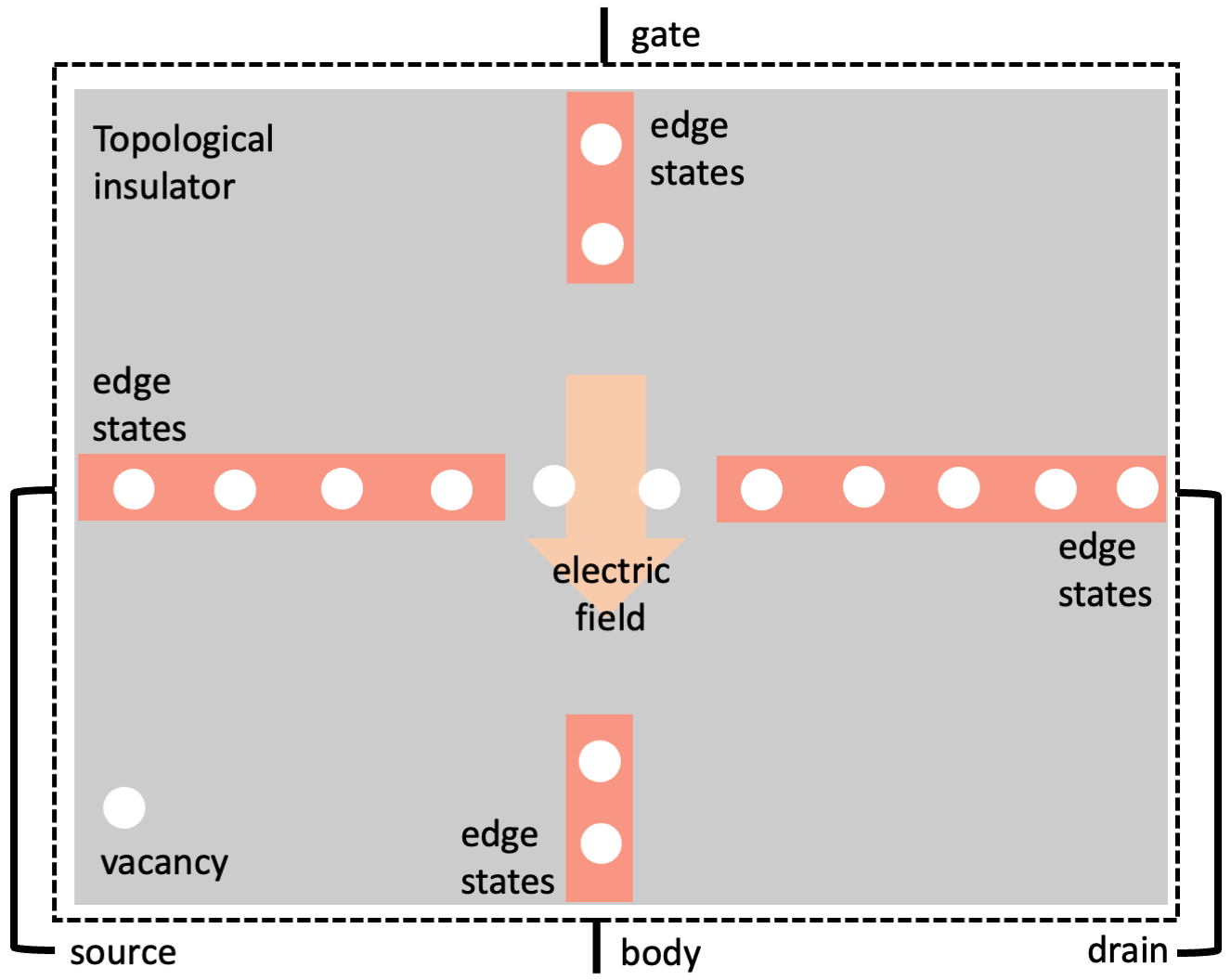}
    \caption{The structure of FET in TIs using vacancies induced edge states. A chain of vacancies can be used as wire connecting source and drain terminals. The gate terminal can also made by vacancies but spacing with a layer of vacancies-free topological insulator. The layer of vacancies-free area allow the field of gate terminal affecting states on the chain of vacancies but avoid breakdown.}
    \label{figstructure}
\end{figure}

Fig.\ref{figstructure} show schematically the structure of junction FET based on vacancies induced edge states in TIs. The conducting channel between source and drain terminals is provided by a chain of vacancies. The resistance of the conducting channel is determined by the parameters of topological material and inter-vacancies distance.\cite{brotherpaper} The gate terminal is also a chain of vacancies perpendicular to the conducting channel between source and drain terminals. However, the gate terminal should keep some distance from the conducting channel for avoiding breakdown between gate terminal and the other two. The distance between gate terminal and the conducting channel should be controlled as zero overlap of their induced edge states. Therefore the region between gate terminal and the conducting channel is insulating with the size of energy gap equal to the energy difference between bulk states and edge states as shown in Fig.\ref{figband}-b. The gate terminal is like a capacitor around conducting channel that produce electric field in between the two legs of gate terminal. The insulating region allow electric field reach the affect the conducting channel but prevents current conduction with it. 

\begin{figure}[ht!]
    \centering
    \includegraphics[width=\linewidth]{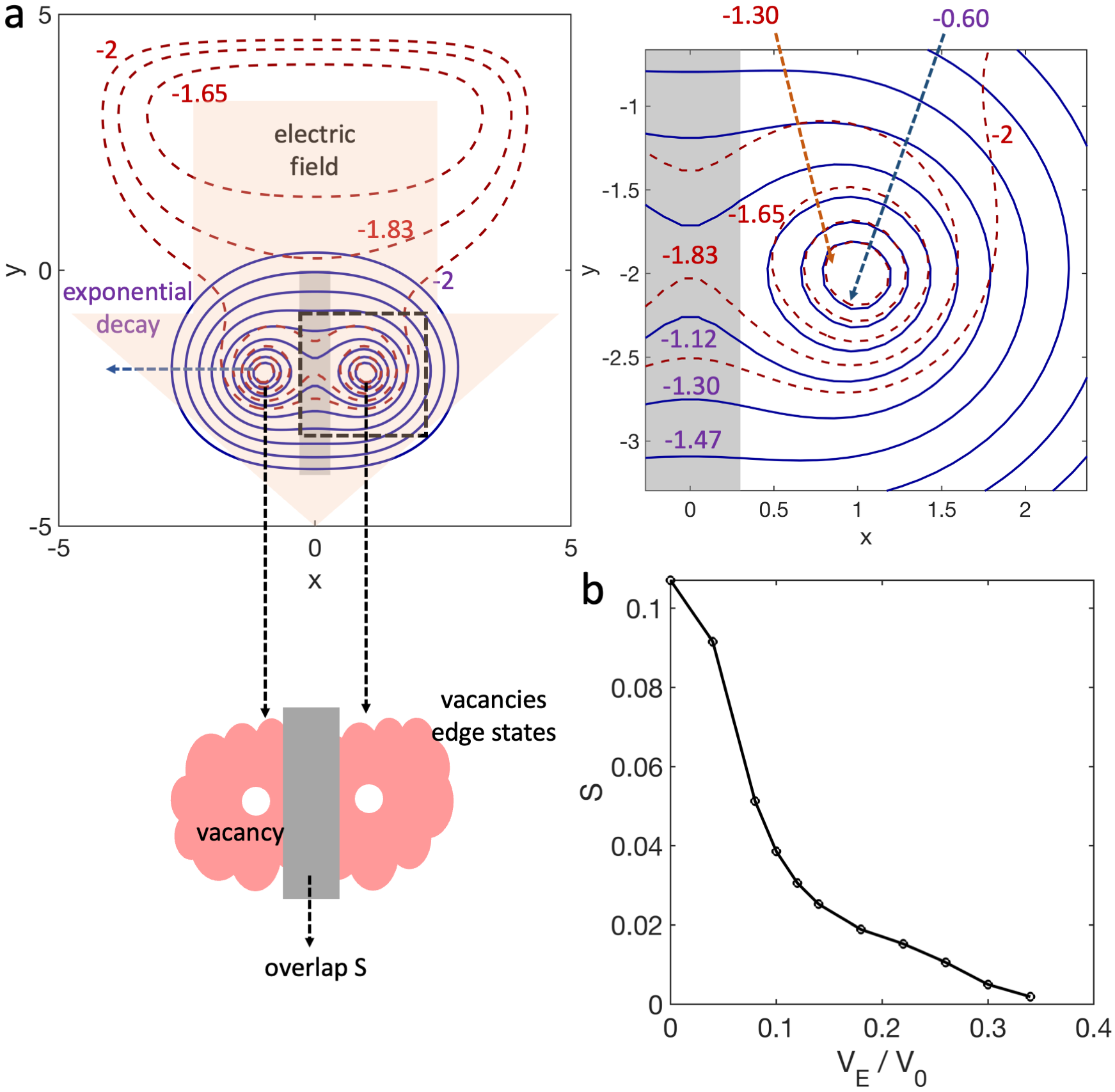}
    \caption{The effects of electric field on the distribution of carriers of vacancies induced edge states. \textbf{a}, contour map of the ground state carriers' distribution of vacancies induced edge states without electric field (blue) and under uniform electric field (red) obtained using density functional theory simulation. The density have been normalized to \(1\), \(\int \rho (x,y) dxdy = 1\). Contour levels are logarithm density, \(\mathrm{log}_{10} \rho\). The area in dashed box is enlarged in right top. For obtaining red dashed ground state in this panel, \(E_y = 0.15\) is used. \textbf{b}, the overlap density in function of relative electric field intensity. The overlap \(S\) is defined as density of carriers under the gray area between two vacancies, \(S = \int_{gray} \rho (x,y) dxdy\), which determines if current are allowed moving from one vacancy to another. The relative electric field intensity is defined as \(V_E / V_0\), \(V_E = E_y L_y\) is the depth of electric potential and \(V_0\) is the depth of vacancy's potential which reflect the robustness of states induced by vacancies. In the simulation, \(L_x = 10\) and \(V_0 = 10\) are used.}
    \label{figefield}
\end{figure}

To clarify the field effect on conducting channel in Fig.\ref{figstructure}, the result of simulations using DFT is shown in Fig.\ref{figefield}. The DFT Hamiltonian functional is,\cite{RevModPhys.87.897,PhysRev.140.A1133}
\begin{equation}
    H[\rho] = -\dfrac{1}{2} \nabla^2 + V_{P} + V_{ext} + V_H [\rho],\label{dfth}
\end{equation}
where the first term represent kinetic energy of electrons, the second term \(V_P\) is the pseudopotential used to simulate density distribution of topological edge states, the third term \(V_{ext} = - E_y \cdot y\) the external potential energy of electrons under electric field applied by gate terminal, and the Hartree potential \(V_H [\rho]\) is the solution of Poisson's equation \(\nabla^2 V_H = -4 \pi \rho\). The pseudopotential,
\begin{equation}
    V_{P}(x, y) =
\begin{cases}
-V_0, & \text{if } \sqrt{(x - x_c)^2 + (y - y_c)^2} \leq R, \\
0, & \text{otherwise},
\end{cases}
\end{equation}
is a square well potential with depth \(V_0\), position \(\boldsymbol{r}_c = (x_c , y_c)\) and radius \(R\) which is the size of vacancy. This pseudopotential is able to reproduce the exponentially decaying ground state density distribution without electric field \(\rho(\boldsymbol{r}) \propto e^{-|\boldsymbol{r}-\boldsymbol{r}_c|/\xi}\) with \(\xi\) the characteristic decay length for topological edge states induced by vacancy,\cite{brotherpaper,PhysRevB.92.195107} as show in Fig.\ref{figefield}-a. The Hamiltonian \(H[\rho]\) in Eq.\eqref{dfth} can be then used for computing ground state density distribution of topological edge states iteratively by solving Kohn-Sham equation,\cite{RevModPhys.87.897,PhysRev.140.A1133}
\begin{equation}
    H[\rho]\psi [\rho] = E[\rho] \psi [\rho],
\end{equation}
until find the ground state distribution \(\rho\) which gives minimum eigen energy. More details about DFT simulation can be found in Supplementary Information, SI-2. 

The Hamiltonian in Eq.\eqref{dfth} do not involve topological phase transition, since topological phase transition is not necessary to reach the open circuit condition. Topological phase transition usually need relatively strong electric field to remove the edge state in principle.\cite{doi:10.1126/science.1256815} However, to open the conductance of a chain of vacancies, it only needs to reducing the local states density, \(\rho\), to weaken the connection between the two nearby vacancies,\cite{brotherpaper} which is same as reducing the quasi Fermi level in to the band gap between vacancies states and normal bulk states as illustrated in Fig.\ref{figfermi}. The Hamiltonian in Eq.\eqref{dfth} is able to show the effect of electric field on ground state distribution of exponentially localized states, and hence is helpful for current analysis.

In Fig.\eqref{figefield}-a, the ground state distribution of vacancies induced edge state around two vacancies under is shown. Blue solid lines and red dashed lines are results under zero external electric field and non-zero electric field respectively. It can be seen that the vacancies induced topological states is fragile under field effect, the density can be reduced around \(10\) times under relatively weak electric field comparing with depth of pseudopotential
\(V_0\), \(E_y L_y / V_0 = 0.15\). The result show that the exponentially decayed topological edge states induced by vacancies can be affected obviously by applying weak external field and as schematically designed in Fig.\ref{figstructure}. Fig.\ref{figefield}-b show the ground state density under gray region in Fig.\ref{figefield}-a as a function of relative field intensity \(V_E = E_y L_y\). The gray region lies in the center between two vacancies. The ground states density under this region is vital for the conductance of current since the effective hopping parameter \(t_m\) in Eq.\eqref{realh} between two vacancies is proportional to the ground density under gray, \(t_m \propto \int_{gray}\rho d\boldsymbol{r}\). When the ground state under gray region is reduced to zero, the effective hopping of electrons between the two vacancies will be totally forbidden and hence lead to open circuit condition on the conducting channel of source and drain terminals shown in Fig.\ref{figstructure}. From the other side, the qausi Fermi energy, \(E_F^* \approx E_F + k_B T \mathrm{ln}(\rho / \rho_{0})\), at the two vacancies will also be decreased due to decreasing of ground state density \(\rho\). When the qausi Fermi energy lies into the gap between edge and bulk states shown in Fig.\ref{figband}, the conducting channel locally is in fact an insulating condition and the circuit between source and drain will be open.

Topological edge states induced by translational symmetry breaking is robust under disorder, even so, the exponential decaying distribution also indicate their fragility under perturbation of external field.\cite{PhysRevLett.123.206801} Form the pseudopotential, it can have a glimpse the fragility is due to edge effect that indicates effective potential behaving as a square potential well. Such fragility provides opportunities to controlling open and closed circuit through field effect based on edges states of TIs, and hence the route of designing FET electronic devices. Inducing that vacancies in \(2D\) materials is able to be realized, the advantages of using vacancies for TIs FET is also obvious. Firstly, the dynamics of vacancies induced edge states are know,\cite{brotherpaper} hence the energy band of conducting line consistent by a chain of vacancies is easy to be controlled. Then, the states induced by vacancies can be affected by relatively weak electric field which is necessary for low power devices. The size of vacancy is in atomic scale in order of \(\sim 0.1\) nanometers, which is much smaller than the typical scale of modern electronic devices using semiconductor. As an example, \(3\) nanometers process of semiconductor manufacturing by Intel, the scale of interconnect pitch is in order of \(\sim 10\) nanometers.\cite{3nm}

Nevertheless, in addition to promising qualitative considerations here, further realistic simulation and experimental confirmation will be needed to answer few vital questions, that including the efficiency of vacancies chain as conducting channel, the stability of gate terminal after applying voltage, and the field effect of conducting vacancies states.

\section{Conclusion}
In this work, a scheme for atomic scale junction FET based on vacancies edge states in TIs is proposed. According to the dynamics of vacancies edge states, the conducting channel between source and drain terminals can be provided by a chain of vacancies, for such a structure its energy bands can have a gap between edge and bulk states. Then the quasi fermi level on the conducting channel can be locally shifted into the gap through field effect applied by gate terminal and hence leads to locally insulating state on the conducting channel and turn off the connection between source and drain terminals. Haldane model and DFT simulation results confirmed band structure and electric  field effect respectively. This work suggest promising application potential of vacancies in TIs for atomic scale logic circuit. 

\section*{Acknowledgments}
The author would like to thank Matteo Baggioli, Weicen Dong and Haifei Qin for illuminating discussions.

\section*{Appendix A: Haldane model of topological insulator}
Haldane model is adopted to simulate the effect of edge and vacancies and their associated excitations in topological insulators, which Hamiltonian in real space reads,\cite{PhysRevLett.61.2015}
\begin{equation}
    H = \sum_{i,j} t_1 c_{i}^{\dagger} c_{j} \delta_{r_{ij},R_n} + \xi t_2 i c_{i}^{\dagger} c_{j} \delta_{r_{ij},R_s},
\end{equation}
In the Hamiltonian, $c_{i}^{\dagger} (c_{i})$ is the creation(annihilation) operator act on a real space site $\boldsymbol{r}_{i}$ with $i$ denotes atom ID, $t_1$ and $t_2$ are the nearest neighbor and the second nearest neighbor hopping parameters. The Delta functions $\delta_{r_{ij},R_n}$ and $\delta_{r_{ij},R_s}$ promise the nearest neighbor and the second nearest neighbor hopping with $R_n$ and $R_s$ the two distances respectively. $\xi = \pm 1$ determined by hopping direction and A(B) lattice sites as in standard Haldane model.\cite{PhysRevLett.61.2015}

To obtain the band structure, Hamiltonian in wave vector space can be constructed by using real space Hamiltonian with Bloch theorem,
\begin{equation}
    H(\boldsymbol{k})_{i,j} = H_{i, j} e^{-i \boldsymbol{k} \cdot (\boldsymbol{r}_j - \boldsymbol{r}_i )}.
\end{equation}
The band structure \(E(\boldsymbol{k})\) is therefore the eigenvalue of the wave vector space Hamiltonian,
\begin{equation}
    H(\boldsymbol{k}) \psi_{\boldsymbol{k}} = E(\boldsymbol{k}) \psi_{\boldsymbol{k}}. 
\end{equation}

\section*{Appendix B: Simulation of vacancies induced edge states under perturbation of additional electric field using density functional theory}

Density functional theory (DFT) is widely used for finding the ground state distribution of electrons for a many body problem with interactions. The main procedure of DFT algorithm is the following. The original many body problem of electrons can be simplified to single particle problem after adiabatic approximation, which assume that the external filed \(V_{ext}\) is independent of electrons' distribution \(\rho(\boldsymbol{r})\), and Hartree approximation, which simplify the many body interactions \(V_H [\rho]\) between electrons into a functional of electron density.\cite{RevModPhys.87.897} The Hamiltonian is therefore also a functional of electron density,
\begin{equation}
    H[\rho] = T[\rho] + V_P + V_{ext} + V_H [\rho],\label{eqsihamiltonian}
\end{equation}
with \(T[\rho]\) the kinetic energy functional. The above Hamiltonian functional can be used further to solve Kohn-Sham equation,\cite{PhysRev.140.A1133}
\begin{equation}
    H[\rho] \psi [\rho] = E[\rho] \psi [\rho]
\end{equation}
for obtaining the minimum eigen energy \(E_m [\rho]\) and corresponding eigen wave function \(\psi_m [\rho]\). The corresponding electron density can be updated to \(\rho_m = |\psi_m [\rho]|^2\). Since the real ground state energy \(E_0 \leq E_m [\rho]\) is the lowest, one can repeat above computation by replacing \(\rho\) by \(\rho_m\) in Eq.\eqref{eqsihamiltonian} and finally solve the ground density distribution self-consistently \(\rho_0 \approx E_m [\rho]\) after iterations.

In practical, the pseudopotential potential \( V_{P} \) consists of two square wells with a depth \( V_0 = 10\) and radius \( R = 0.2\), defined as,
\[
V_{P}(x, y) =
\begin{cases}
-V_0, & \text{if } \sqrt{(x - x_c)^2 + (y - y_c)^2} \leq R, \\
0, & \text{otherwise}.
\end{cases}
\]
where \( (x_c, y_c) \) are the centers of the wells, are used for simulate the effect of vacancy on its around edge states with exponential decay. The size of simulation area is a \(10 \times 10\) box. The Hartree potential \( V_H \) is computed using a Gaussian kernel to approximate the Coulomb interaction,
\[
V_H(x, y) = V_{H0} \cdot \frac{\rho \star G(x, y)}{\max(V_H)},
\]
where \(\star\) denotes convolution computation and \( G(x, y) \) is the Gaussian kernel,
\[
G(x, y) = \frac{1}{2 \pi \sigma^2} \exp\left(-\frac{x^2 + y^2}{2 \sigma^2}\right),
\]
with \(\sigma = 0.1\) the standard deviation. An external uniform electric field \( E_y \) is applied, adding a linear potential along \(y\) direction,
\[
V_{ext}(x, y) = -E_y \cdot y,
\]
is also added to simulate the effects of external field on topological states around vacancies.

The kohn-Sham equation is solved using the finite difference method. The Hamiltonian \( H \) is given by,
\[
H = -\frac{1}{2} \nabla^2 + V_{\text{total}}(x, y),
\]
where \( V_{\text{total}}(x, y) = V_{P}(x, y) + V_H(x, y) + V_{ext}(x, y) \). The Laplacian operator \( \nabla^2 \) is discretized using finite difference matrices,
\[
\nabla^2 \approx \frac{\partial^2}{\partial x^2} + \frac{\partial^2}{\partial y^2},
\]
where,
\[
\frac{\partial^2}{\partial x^2} = \frac{1}{\Delta x^2}
\begin{bmatrix}
-2 & 1 & 0 & \cdots & 0 \\
1 & -2 & 1 & \cdots & 0 \\
0 & 1 & -2 & \cdots & 0 \\
\vdots & \vdots & \vdots & \ddots & \vdots \\
0 & 0 & 0 & \cdots & -2
\end{bmatrix}.
\]

The eigenvalue problem is solved by diagonalizing Hamiltonian,
\[
H \psi = E \psi,
\]
where \( \psi \) is the wavefunction and \( E \) is the energy eigenvalue. To obtained the ground state density distribution, the self-consistent loop iteratively updates the electron density \( \rho_{new}(x, y) = (1-m) \rho_{old}(x, y) + m |\langle (x,y)|\psi\rangle|^2 \) until convergence within a small tolerance. Here \(m = 0.1\) is the updating rate control the speed of self-consistent iteration.

\end{document}